# Vacancy defects and monopole dynamics in oxygen-deficient pyrochlores

G. Sala[1], M. J. Gutmann[2], D. Prabhakaran[3], D. Pomaranski[4,5], C. Mitchelitis[4,5], J. B. Kycia[4,5], D. G. Porter[1], C. Castelnovo[6] and J. P. Goff[1*]

[1]Department of Physics, Royal Holloway, University of London, Egham, TW20 0EX, UK
[2]ISIS Facility, Rutherford Appleton Laboratory, Chilton, Didcot OX11 0QX, UK
[3]Department of Physics, University of Oxford, Oxford OX1 3PU, UK
[4]Department of Physics and Astronomy and Guelph-Waterloo Physics Institute, University of Waterloo, Waterloo, Ontario N2L3G1, Canada
[5]Institute for Quantum Computing, University of Waterloo, Waterloo, Ontario N2L 3G1, Canada
[6]Theory of Condensed Matter group, Cavendish Laboratory, University of Cambridge, Cambridge CB3 0HE, UK

**The idea of magnetic monopoles in spin ice has enjoyed much success at intermediate temperatures, but at low temperatures a description in terms of monopole dynamics alone is insufficient. Recently, numerical simulations were used to argue that magnetic impurities account for this discrepancy by introducing a magnetic equivalent of residual resistance in the system. Here we propose that oxygen deficiency is the leading cause of magnetic impurities in as-grown samples, and we determine the defect structure and magnetism in $Y_2Ti_2O_{7-\delta}$ using diffuse neutron scattering and magnetization measurements. These defects are eliminated by oxygen annealing. The introduction of oxygen vacancies causes $Ti^{4+}$ to transform to magnetic $Ti^{3+}$ with quenched orbital magnetism, but the concentration is anomalously low. In the spin-ice material $Dy_2Ti_2O_7$ we find that the same oxygen-vacancy defects suppress moments on neighbouring rare-earth sites, and that these magnetic distortions dramatically slow down the long-time monopole dynamics at sub-Kelvin temperatures.**

The frustrated geometry of the pyrochlore lattice of corner-sharing tetrahedra, and the crystal electric field effects that constrain the rare-earth magnetic moments to point along the tetrahedron axes, are responsible in great part for the behaviour of the spin-ice materials[1]. The leading effect is the appearance of "2in-2out" ice rules[2-5], analogous to the proton arrangement in water ice[6]. These local constraints give rise to an intriguing display of topological properties[7]. The frustrated geometry in these systems provides the key to a longstanding problem in theoretical physics: how to stabilize fractional excitations in three dimensions. The proposal that excitations in spin ice fractionalize into de-confined magnetic monopoles interacting via the magnetic Coulomb interaction[8] has led to intense experimental activity in recent years. In the case of diffuse neutron scattering it was possible to observe the "Dirac strings" that trace the random walk of monopoles[9], and to infer the presence of monopoles from the broadening of "pinch-point" features[10-11]. Magnetic currents have been observed in a magnetic field[12], and magnetic relaxation measurements have been interpreted in terms of the diffusive dynamics of free monopoles[13-15].

However, recent experiments conducted at sub-Kelvin temperatures show that magnetization dynamics in spin ice samples occurs on far longer time scales than one could explain using straightforward monopole hydrodynamics, even accounting for Coulomb interactions[16-18]. In an attempt to explain this discrepancy, magnetic impurities were shown to be capable of dramatically reducing the flow of magnetic monopoles, similarly to electrical conductors in which local impurities can decrease the electrical conductivity[19]. To date, magnetic impurities in spin ice have been modelled based on the assumption that they resemble "stuffed spin ice"[20]. The determination of the nature of the defects in as-grown samples, whether they comprise substitutions or vacancies, the extent of the distortion of the surrounding lattice, and the effects on the magnetic properties, has become a pressing issue. Understanding these defects is crucial for experiments directed at single monopole detection, the observation of monopole currents, and the design of potential spin-ice devices.

We answer this question by first investigating defect structures in oxygen-depleted pyrochlores and then studying how the oxygen vacancies affect the magnetism of spin-ice materials. The absence of a large-moment rare-earth ion in $Y_2Ti_2O_{7-\delta}$ allows us to focus on the structure and on the magnetism of the Ti sites. The structure of the defect clusters in oxygen-deficient $Y_2Ti_2O_{7-\delta}$ was determined using diffuse neutron scattering, which is particularly sensitive to vacancies and displacements of oxygen ions. We also find that oxygen vacancies are the dominant defects in as-grown "stoichiometric" samples, rather than the stuffing of Ti sites with Y. The removal of $O^{2-}$ ions results in the replacement of $Ti^{4+}$ ions by $Ti^{3+}$, and this leads to the introduction of magnetic moments on the Ti sites. Our results imply that $Ti^{3+}$ ions are far from being the leading magnetic perturbation in spin-ice materials. We relate our results to $Dy_2Ti_2O_{7-\delta}$ and we find that oxygen vacancies change the nature of the magnetism on the surrounding four $Dy^{3+}$ ions. We find that these magnetic impurities dramatically change the low-temperature monopole dynamics.

All of our x-ray diffraction data for the black oxygen-depleted $Y_2Ti_2O_{7-\delta}$, the yellow as-grown and transparent annealed $Y_2Ti_2O_7$ refine in the cubic pyrochlore structure, space group $Fd\bar{3}m$. Y and Ti ions are located on pyrochlore lattices, and there are two inequivalent O sites: O(1) located at the centre of the Y tetrahedra, and O(2) filling interstitial regions. Refinement of x-ray diffraction data reveals equal concentrations on the Y and Ti sites to within 2%, see Table 1, so that stuffing of Ti sites by Y is minimal. The stoichiometry of the oxygen-depleted sample determined by thermo-gravimetric analysis (TGA) was $Y_2Ti_2O_{6.79}$, and the diffraction data are consistent with these values. The vacancies were found to be mainly on the O(1) sites, in agreement with the literature that reports that the O(1) are not as strongly bound to the lattice as the O(2) ions[21,22]. The length of the unit cell increases monotonically with decreasing oxygen concentration, as expected.

Diffuse neutron scattering is very sensitive to departures from ideal stoichiometry. $Y_2Ti_2O_{6.79}$ exhibits strong, highly anisotropic diffuse scattering throughout reciprocal space.

Figure 1 shows the diffuse scattering from $Y_2Ti_2O_{6.79}$ in the ($hk$7) plane, which is particularly informative. The diffuse scattering in this plane comprises a distinctive pattern of four rods that create a cross at the centre of the plane and four arcs that link the rods. Figure 2 shows that the as-grown sample has qualitatively very similar, but lower intensity scattering. Thus we conclude that the oxygen-depleted and as-grown samples have very similar defect structures. The fact that no diffuse scattering was detected for the annealed sample shows that it has very few defects, and that it is possible to obtain a sample much closer to ideal stoichiometry by annealing in oxygen. It also suggests that oxygen vacancies are the main defects in the as-grown sample, and seems to rule out the possibility of stuffing of the Ti lattice by Y ions on the same scale as oxygen vacancies. On the basis of the diffuse scattering intensities, we estimate compositions $Y_2Ti_2O_{6.97}$ in $Y_2Ti_2O_{7.00}$ for the as-grown and annealed samples.

We developed a Monte Carlo code[23,24] that is able to reproduce qualitatively the main features of the diffuse scattering. We are only able to reproduce the observed crosses when there are relatively large relaxations of Y ions away from isolated O(1) vacancies along <111> directions. The physical origin of this is the Coulomb repulsion between the O(1) vacancies and the Y ions that leads to the expansion of the Y cage. We replace two $Ti^{4+}$ in neighbouring tetrahedra by $Ti^{3+}$ for charge compensation, and move neighbouring O(2) towards $Ti^{4+}$ sites so that $Ti^{3+}$-O and $Ti^{4+}$-O bond lengths agree with those in the literature. The smaller displacements of the surrounding ions were simulated using the "balls and springs" model[25], and this successfully reproduces the four arcs that link the branches of the cross, see Fig. 1(b). Neighbouring O(1) are pushed away along <111> directions by Y ions. We found that there is no evidence for correlations between the O(1) vacancies. The defect structure around an O(1) vacancy is shown in Fig. 1(a). Replacement of the $Ti^{3+}$ ions by $Y^{3+}$ ions would give the stuffed spin ice $Y_2(Ti_{2-x}Y_x)O_{7-x/2}$. Although the diffuse neutron scattering is consistent with a low level of stuffing, the x-ray diffraction effectively rules it out. This is in contrast to $Yb_2Ti_2O_7$, where the smaller Yb ions are found to substitute for Ti at the percent level[26]. Even where stuffing does occur, it is highly likely that charge compensating O(1) vacancies are important defects.

For $Y_2Ti_2O_{7-\delta}$, the removal of $O^{2-}$ ions changes the oxidation state of Ti ions in order to preserve charge neutrality. $Ti^{3+}$ ions are magnetic and for modest concentrations of vacancies $Y_2Ti_2O_{7-\delta}$ becomes paramagnetic, as the leading dipolar interaction between them is negligible at the temperatures of interest. Consistently, we did not detect any magnetic diffuse scattering from $Y_2Ti_2O_{6.79}$ using unpolarised neutrons on SXD down to 0.3K. The SQUID data presented in Fig. 3 are fitted with a Brillouin function with S = ½, and this suggests that the orbital moment is quenched. However, the concentration of defects $\delta = 0.023$ is much lower than the values obtained using structural refinement or TGA. One possible explanation is that neighbouring $Ti^{3+}$ form strong antiferromagnetic bonds and the observed signal is from isolated $Ti^{3+}$ ions. Another possibility is that a sizable proportion of the Ti ions form $Ti^{2+}$, which is expected to have a singlet ground state. It is even possible that partial charge compensation may be achieved through trapped electrons on vacancies, via the so-called F-centres often responsible for colour in this class of material[27]. Incidentally, the presence of almost-free $Ti^{3+}$ ions may be the rapidly fluctuating magnetic impurities required to understand the NMR relaxation at low temperature[28].

Annealing as-grown $Dy_2Ti_2O_7$ in oxygen leads to defect-free transparent yellow crystals and, therefore, we conclude that the dominant defects in this case are also oxygen vacancies. Our diffuse neutron scattering studies show that oxygen vacancies are located on the O(1) sites for $Dy_2Ti_2O_7$, see Fig. 4(a). The oxygen concentration measured using TGA is $\delta = 0.02$ for the as-grown sample. In Fig. 4(b) we compare the static magnetic susceptibility of an as-grown sample before and after annealing in oxygen. We are able to clearly resolve a reduction in saturation magnetization that implies a reduced moment on the defective Dy sites. Our crystal electric field calculations show that $Dy^{3+}$ ions have a reduced moment in the presence of an O(1) vacancy, and the anisotropy changes from easy axis along <111> for the stoichiometric compound to easy plane perpendicular to the local <111>, see section C of the Supplementary Information.

AC susceptibility measurements were conducted on an as-grown and annealed $Dy_2Ti_2O_7$ crystal at the University of Waterloo using a SQUID Susceptometer on a dilution refrigerator[16]. These samples are both needle-shaped to reduce the demagnetization correction, they have the same crystallographic orientation, and the results were obtained at the same temperature, $T = 800$ mK. The measured imaginary portion of the AC susceptibility, $\chi''(\omega)$, was transformed to the dynamic correlation function $C(t) = <M(0)M(t)>$, where $M(t)$ is the time-dependent magnetization of the sample[29]. The dynamic correlation function results are presented in Fig. 4(c), where they are compared with previous results on a different, non-annealed sample of $Dy_2Ti_2O_7$, at $T = 800$ mK[19]. The very slow long-time tail in $C(t)$ associated with magnetic defects is observed for the as-grown sample, but it is entirely suppressed for the annealed sample. The as-grown sample characteristics match closely with the previous results. Revell *et al.* attributed the long-time tail as being a result of interactions of monopoles with magnetic impurities from a slight level of stuffed sites (substitution of Dy for Ti)[19]. The fact that annealing in oxygen eliminates the long-time tale suggests that the magnetic impurity sites in the as-grown sample results from oxygen vacancies. The stretched exponentials seen in the correlation function at early times for all samples were very similar.

We have conducted preliminary investigations of the effects of the magnetic defects introduced by isolated oxygen vacancies on the monopole dynamics. Figure 5(a) shows a tetrahedron with 4 easy-plane spins surrounding an O(1) vacancy, as suggested by our crystal electric field calculations. The 4 neighbouring tetrahedra have, therefore, 3 easy-axis and 1 easy-plane spin each. The concept of ice rules is no longer valid for any of these defective tetrahedra. On the other hand, farther tetrahedra are minimally perturbed and are expected to behave as in stoichiometric spin ice. Let us consider what happens to the system when a monopole hops from one of the farther stoichiometric tetrahedra onto one of the tetrahedra directly affected by the vacancy, as illustrated in the figure, where an adjacent tetrahedron is shown initially in an excited 3out-1in monopole state. Flipping the spin that joins the stoichiometric tetrahedron to a defective tetrahedron results in the monopole hopping onto one of the 3-easy-axis, 1-easy-plane tetrahedra, see Fig. 5(b). The direction of the 4 easy-plane spins relaxes to minimise the energy of the system given the orientation of the surrounding easy-axis spins, and this is responsible for a reduction in the energy of the system. Pictorially, the monopole charge has "delocalised" over the defective tetrahedra and one can no longer identify a specific tetrahedron where the monopole resides.

In section D of the Supplementary Information we have calculated the energy changes for all configurations of this cluster of spins, for exchange and dipolar interactions truncated at nearest-neighbour distance, with various working assumptions on the exchange interaction involving easy-plane spins. In all cases, we obtained a broad distribution of energies down to large negative values, large enough to be comparable to the bare energy cost of an isolated monopole. This means that a monopole coming into contact with a vacancy cluster can become strongly pinned to it. It is remarkable to notice the similarity in the way that monopoles interact with these oxygen vacancies compared to the static stuffed moments discussed by Revell *et al*[19]. Depending on the direction of approach, the magnetic defects can either attract or repel a monopole. Since the static magnetic moments introduced in the simulations were able to explain the long-time tail in the magnetic relaxation of the system[19], one can reasonably expect that local magnetic distortions introduced by oxygen vacancies can lead to similar long-time tails. The experimental results in Fig. 4(c) demonstrate that this is the case.

One must further notice that, from the very same energetic arguments presented above, these magnetic defects can also act as nucleation centres for monopoles. Hence they may be responsible for an increase in monopole density in thermodynamic equilibrium. In turn, a heightened monopole density means faster (linear-response) dynamics[14]. This is precisely the opposite effect of a pinning centre. Understanding the interplay between these two effects requires careful studies that are beyond the scope of the present article. One might nonetheless speculate that the speedup due to an increase in monopole density is more likely to dominate at short time scales (close to equilibrium), whereas pinning effects become important when the system is driven far from equilibrium at long times. It is intriguing to note that this expectation appears to match the qualitative differences between as-grown and annealed behaviour of $C(t)$ in Fig.4(c), whereby the removal of oxygen vacancies is seen to simultaneously suppress the long-time tail *and* slow down the short-time magnetic relaxation at the same time.

We have clearly demonstrated the important role of oxygen vacancies in the monopole physics at low temperature. Further studies are required to understand why the magnetic response of the Ti sites is suppressed. The magnetism on the rare-earth sites makes a substantial contribution to the monopole dynamics, but it would be of great interest to explore further the effects on the system. Our results make it apparent that the density of oxygen vacancies in spin ice samples is of key importance in the interpretation and understanding of relaxation and response properties, and far from equilibrium behaviour in general. Many experimental results in this direction are already available[13,17-19,30], and in some cases it may be that further work is needed to distinguish the effects of oxygen vacancies from the stoichiometric behaviour.

Moreover, theoretical work to date has largely focused on stoichiometric models[14-15,28,31-34], and it will be interesting to see which new phenomena may emerge when a tuneable density of impurities is introduced via oxygen depletion.

**Methods**

Single crystals of $Y_2Ti_2O_{7-\delta}$ were grown at the Clarendon Laboratory by the floating zone technique. A few pieces of single crystal from the same batch as the as-grown sample were annealed in $O_2$ at a flow rate of 50 mil min$^{-1}$ and a temperature of 1200 °C for 2 days. The as-grown, nominally stoichiometric samples are yellow, and annealing in oxygen produces transparent samples. The oxygen-depleted samples were grown and annealed in flowing mixed gas of hydrogen and argon, and this leads to crystals that are black. The $Dy_2Ti_2O_{7-\delta}$ crystal growth is described in Ref [35]. The average structure was determined using x-ray diffraction at Royal Holloway, and the defect structures were studied by diffuse neutron scattering using the single-crystal diffractometer (SXD) at the ISIS pulsed neutron source at the Rutherford Appleton Laboratory. SXD combines the white-beam Laue technique with area detectors covering a solid angle of $2\pi$ steradians, allowing comprehensive diffraction and diffuse scattering data sets to be collected. Samples were mounted on aluminium pins and cooled to 5K using a closed-cycle helium refrigerator in order to suppress the phonon contribution to the diffuse scattering. A typical data set required four orientations to be collected for 4 hours per orientation. Data were corrected for incident flux using a null scattering V/Nb sphere. These data were then combined to a volume of reciprocal space and sliced to obtain single planar and linear cuts. The DC magnetization was measured using a Quantum Design 7T SQUID magnetometer at the Clarendon Laboratory, and the AC magnetization was measured using a SQUID in a dilution refrigerator at the University of Waterloo. The as-grown $Dy_2Ti_2O_7$ crystal had dimensions $1.0\times1.0\times4.0$mm$^3$, and the annealed $Dy_2Ti_2O_7$ crystal had dimensions $1.0\times0.32\times4.0$mm$^3$, where the long axis was directed along the magnetic field. The data from Revell *et al*. was obtained on a different, non-annealed sample of $Dy_2Ti_2O_7$ with dimensions $1.0\times1.0\times3.9$mm$^3$.

For the Monte Carlo code, a crystal comprising $64 \times 64 \times 64$ unit cells was generated from the average structure obtained from the refinement of the diffraction data. From a statistical perspective, the use of a large supercell helps us to average over the disorder in the system (self average), and suppresses the background noise. O(1) ions are removed at random until we obtain the depletion concentration. Large displacements of surrounding Y ions and smaller displacements of O(2) ions next to $Ti^{3+}$ ions are introduced by hand. The distortion of the surrounding lattice is simulated using the "balls and spring" model in which hard spheres are connected to neighbouring ions by springs, and the simulation randomly displaces ions in order to minimize the elastic energy. Further details of the model are presented in sections A and B of the Supplementary Information. The crystal electric field was calculated using a point-charge model[36], and the monopole dynamics was investigated using cluster calculations, see sections C and D of the Supplementary Information for further details. The magnetization data were modelled using a Brillouin function.

## Acknowledgements

We thank M. Jura and T. Lehner for their help, and B. Gaulin, M. T. Hutchings, R. Moessner, S. L. Sondi, S. Dutton and U.Karahasanovic for helpful discussions. We acknowledge support from the South East Physics Network and the Hubbard Theory Consortium, and we are grateful for the financial support and hospitality of ISIS. This work was supported in part by EPSRC grants EP/G049394/1 and EP/K028960/1, NSERC, the Helmholtz Virtual Institute "New States of Matter and Their Excitations", and the EPSRC NetworkPlus on "Emergence and Physics far from Equilibrium.

## Author contributions

J.P.G., C.C. and D.Pr. designed the research. The neutron measurements were performed by G.S., M.J.G. and J.P.G., the x-ray diffraction was performed by D.G.P. The crystals were grown by D.Pr. The DC magnetisation was measured by D.Pr. and G.S. and the magnetic relaxation measurements were performed by D.Po., C.M. and J.B.K. Theoretical modelling and simulation was by G.S., M.J.G. and C.C. The manuscript was drafted by J.P.G., C.C. and G.S. and all authors participated in the writing and review of the final draft.

## Additional information

Supplementary Information accompanies this paper at http://www.nature.com/nmat/index.html.

|  | **Depleted** | **As-grown** | **Annealed** |
|---|---|---|---|
| Colour | Black | Yellow | Transparent |
| Space group | $Fd\bar{3}m$ | $Fd\bar{3}m$ | $Fd\bar{3}m$ |
| Lattice parameter | 10.123(3)Å | 10.111(3)Å | 10.102(2)Å |
| Y | 0.992(9) | 1.016(5) | 1.01(1) |
| Ti | 1 | 1 | 1 |
| O(1) | 0.88(2) | 1.01(2) | 0.97(3) |
| O(2) | 0.96(2) | 1.06(1) | 1.10(2) |
| $R_w$ | 9.38 | 6.58 | 11.71 |

Table 1. Refinement of the average structure of $Y_2Ti_2O_{7-\delta}$ from the x-ray structure factors measured at $T = 300K$. The Y and Ti occupancies are equal within experimental uncertainty, and the oxygen vacancies in the oxygen-depleted sample are primarily located on O(1) sites.

**Figure captions**

**Fig. 1. Oxygen-vacancy defect structure in $Y_2Ti_2O_{7-\delta}$.** (a) Schematic diagram of O(1) vacancies and the associated distortion of the surrounding lattice, with displacements indicated by green arrows. The four neighbouring Y ions relax away from the vacancy along <111> due to Coulomb repulsion, and they push neighbouring O(1) ions in the same direction. Two $Ti^{4+}$ ions transform to $Ti^{3+}$ to preserve charge neutrality, and nearest neighbour O(2) ions move toward $Ti^{4+}$ ions to give the correct bond lengths. (b) The diffuse neutron scattering in the ($hk$7) plane from $Y_2Ti_2O_{6.79}$ measured at $T = 5$ K (upper half) compared to the Monte Carlo simulation (lower half). The relatively large displacement of the Y ions is needed to obtain the cross, and the O(2) movement is required to reproduce the arcs.

**Fig. 2. Composition dependence of the diffuse scattering.** (a) Cuts through the diffuse scattering from oxygen-depleted (red), as-grown (green) and annealed (blue) $Y_2Ti_2O_{7-\delta}$ along the line highlighted in white in (b). The diffuse neutron scattering in the ($hk$7) plane from as-grown (b) and annealed (c) $Y_2Ti_2O_7$ measured at $T = 5$ K. The diffuse scattering from the as-grown sample closely resembles that from oxygen-depleted $Y_2Ti_2O_{6.79}$ in Fig. 1(b). There is no diffuse scattering from the sample annealed in oxygen.

**Fig. 3. Magnetism on the Ti ions in $Y_2Ti_2O_{7-\delta}$.** (a) The DC magnetization as a function of field in the [111] direction at $T = 2K$ (a) for oxygen depleted $Y_2Ti_2O_{6.79}$ and (b) for as-grown and annealed samples. The experimental data are shown as red circles, and the lines show the fits of Brillouin functions for the Hund's-rule value J = 3/2 (blue) and the orbitally-quenched value J = ½ (green). The saturation levels for the as-grown and annealed $Y_2Ti_2O_7$ samples are a few thousandths of a percent of the value for $Dy_2Ti_2O_7$, suggesting that the induced Ti magnetism does not significantly affect the monopole dynamics in spin ice.

**Fig. 4. Defect structure and magnetism on the Dy ions in $Dy_2Ti_2O_{7-\delta}$.** (a) Diffuse neutron scattering from $Dy_2Ti_2O_{7-\delta}$ reveals the same defect structure as Fig. 1(a) with Y ions replaced by Dy ions. (b) DC magnetization as a function of field in the [100] direction at $T = 2K$ for an as-grown sample of $Dy_2Ti_2O_7$ before and after annealing in oxygen. The presence of oxygen vacancies reduces the saturation magnetization. (c) The dynamic correlation functions from the AC susceptibility. The as-grown sample exhibits the long-time tail seen previously and attributed to magnetic defects[19], but this tail is entirely suppressed by annealing in oxygen. The inset shows Log[$C(t)$] versus time on a log-log scale, where a linear regime identifies stretched exponential behaviour.

**Fig. 5. Monopole trapping by oxygen vacancies in $Dy_2Ti_2O_7$.** An O(1) vacancy is surrounded by four easy-plane spins (green-tip arrows) that are free to rotate in the plane perpendicular to the local [111] directions (green semi-transparent discs). Easy-axis spins are shown with red-tip arrows. The nearest-neighbour tetrahedra are in 2in–1out and 2out–1in easy-axis spin configurations. (a) A 3out-1in monopole is located in the next-nearest-neighbour tetrahedron. (b)

When the yellow arrow is flipped, the monopole hops to the nearest-neighbour tetrahedron, which changes from 2in–1out to 2out–1in. The easy-plane spins are able to relax in response to the change in orientation of the neighbouring easy-axis spins. The energy of (b) is substantially lower than (a), strongly pinning the monopole to the vacancy.

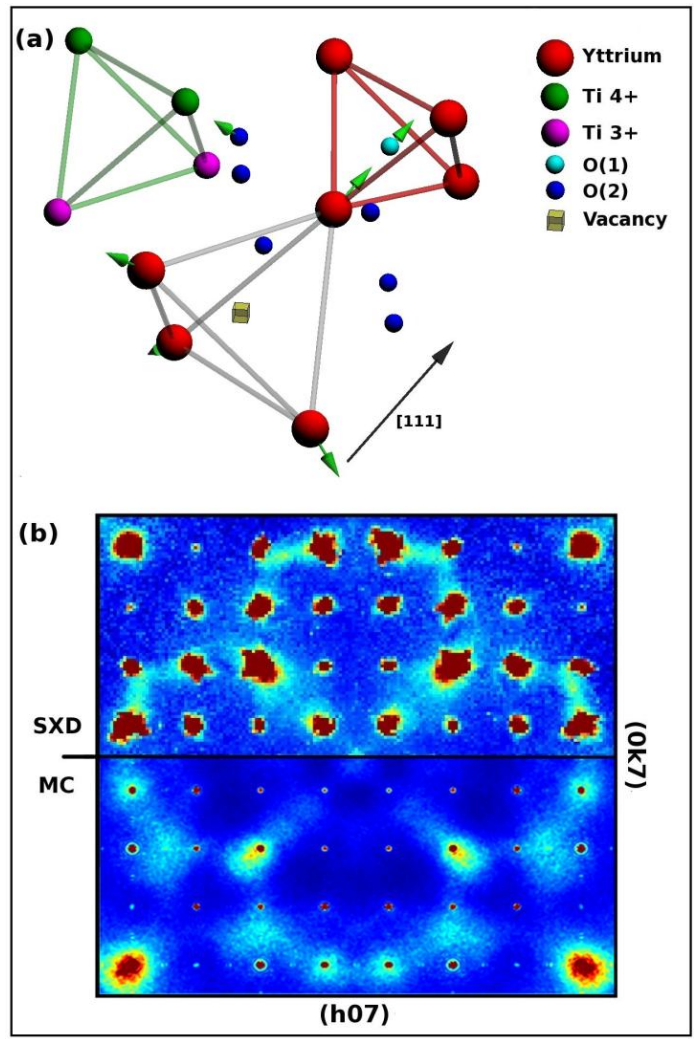

Fig. 1

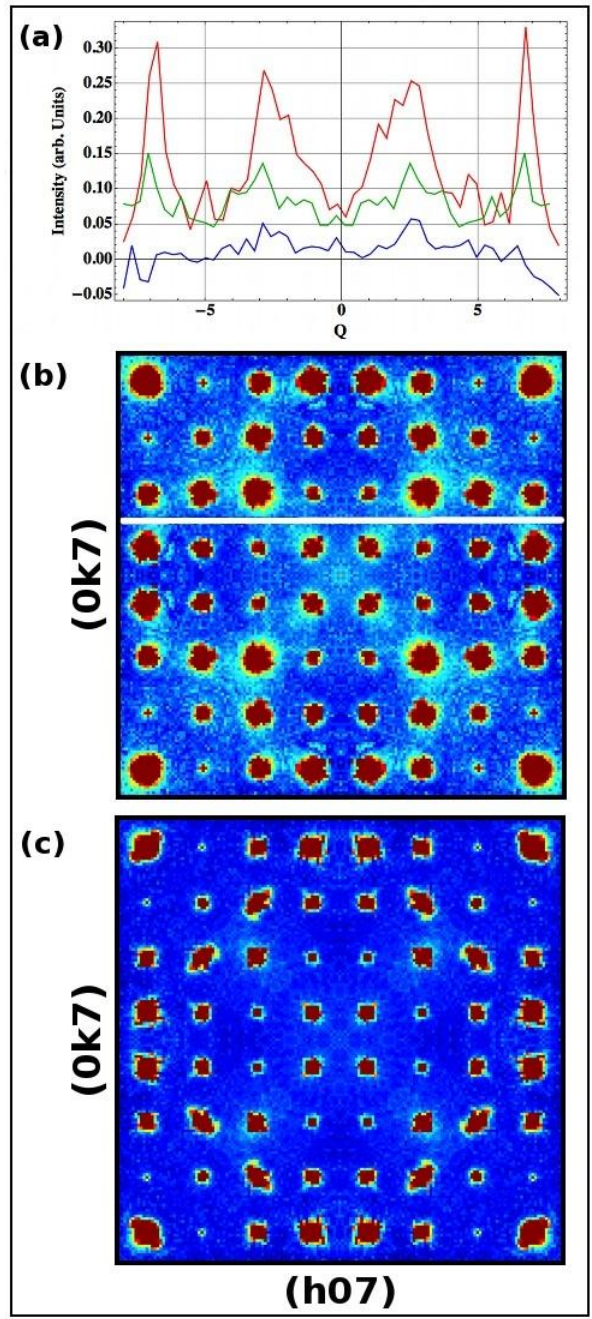

Fig. 2

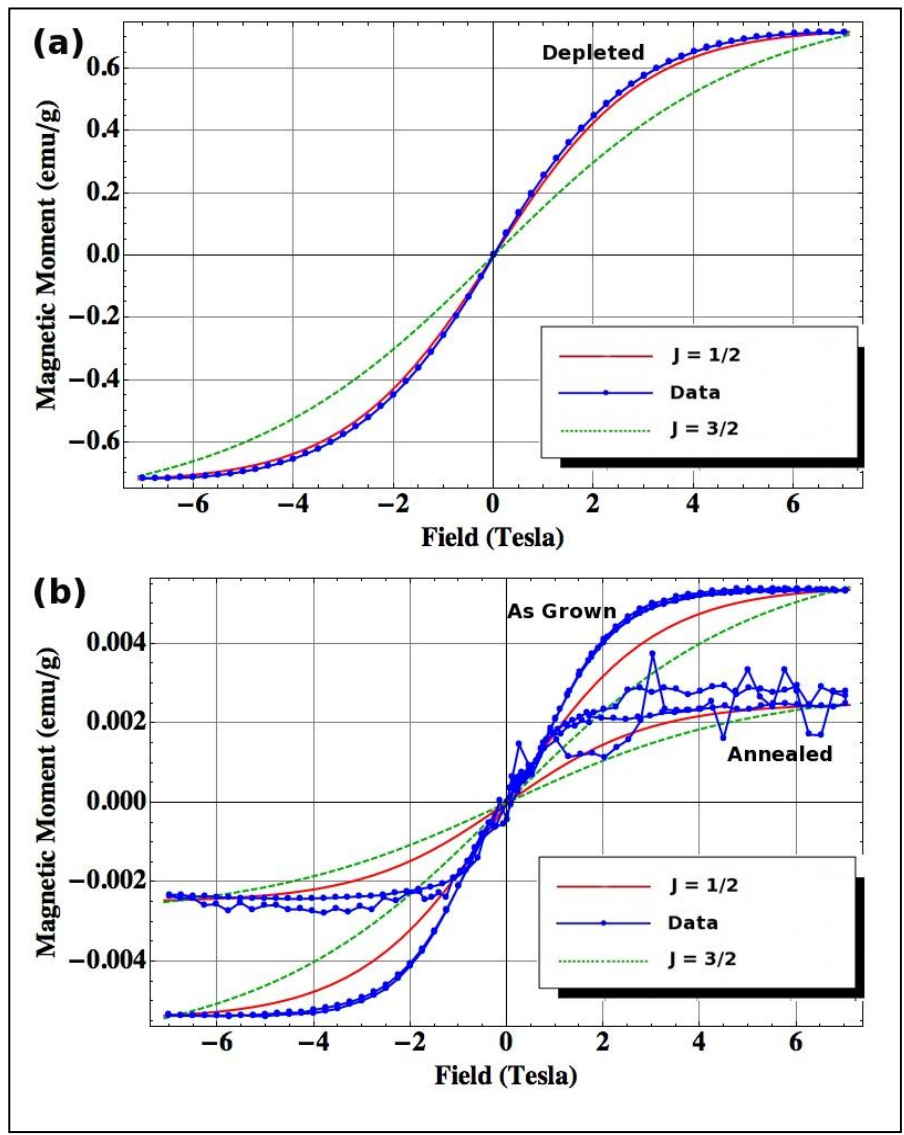

Fig. 3

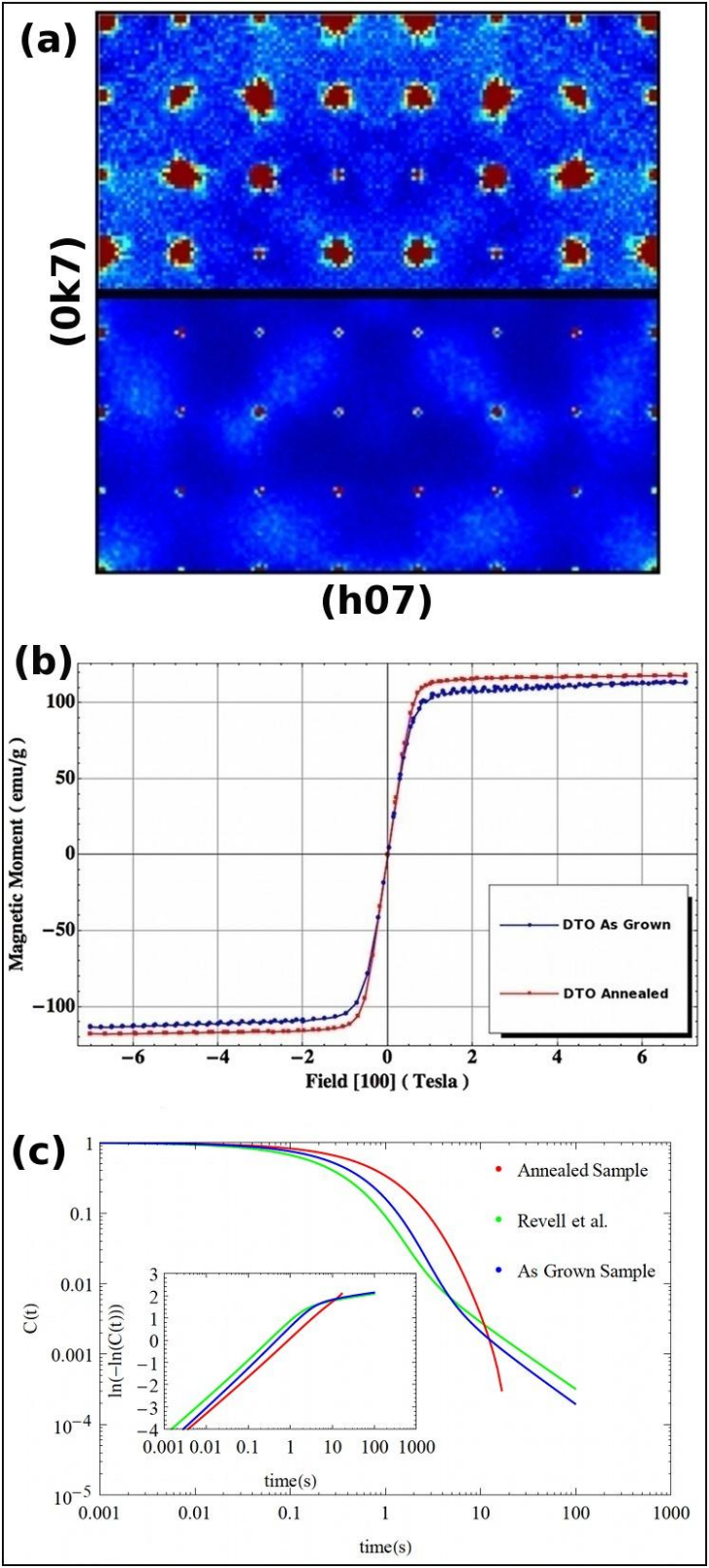

Fig. 4

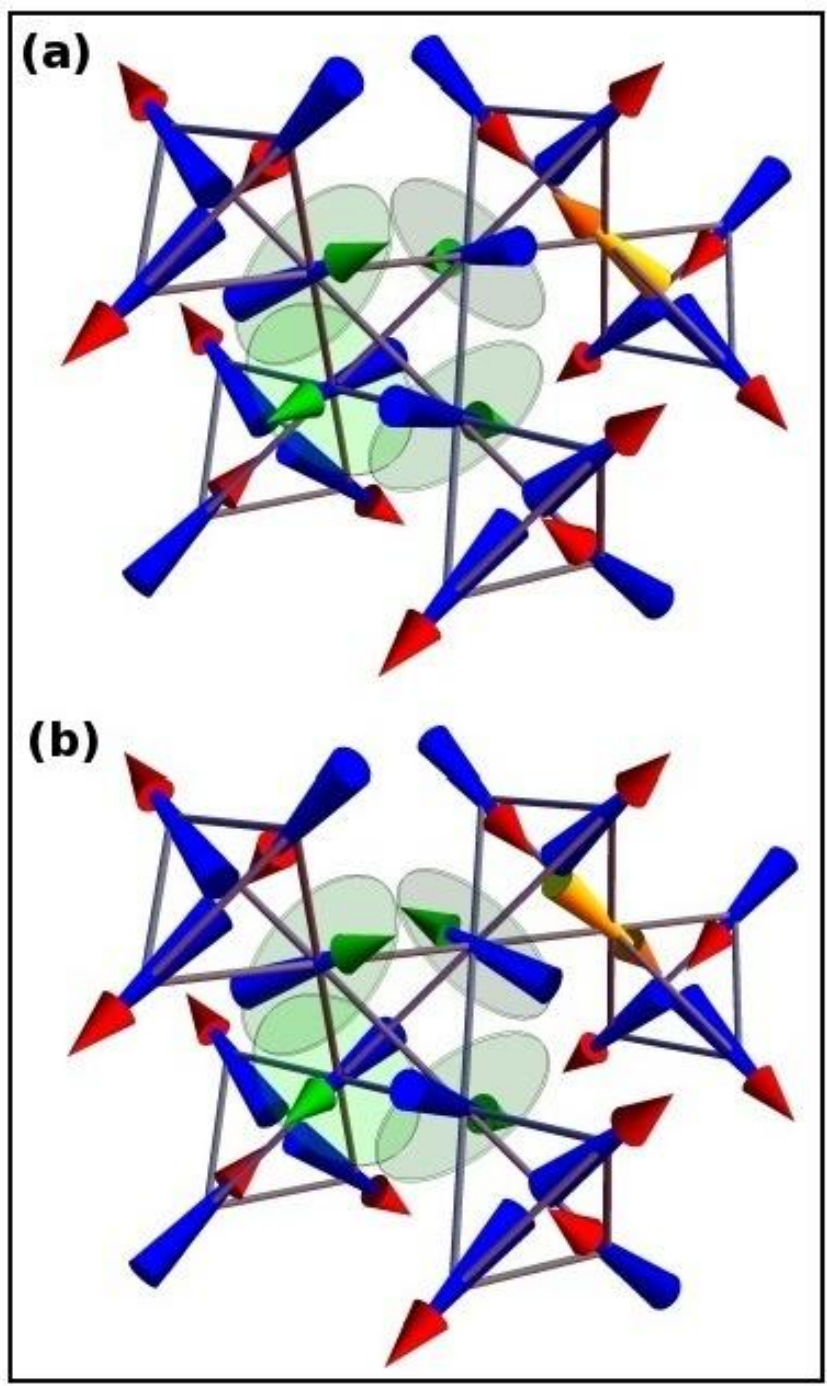

Fig. 5